\documentclass[namedreferences]{kluwer}
\usepackage{epsfig}
\usepackage{psfig}
\def\gapprox{\;\rlap{\lower 2.5pt            
 \hbox{$\sim$}}\raise 1.5pt\hbox{$>$}\;}       
\def\lapprox{\;\rlap{\lower 2.5pt            
 \hbox{$\sim$}}\raise 1.5pt\hbox{$<$}\;} 
\usepackage{graphics}
\usepackage{latexsym}
\begin{document}
\begin{article}
\begin{opening}
\title{Energy budget and imaging spectroscopy of a compact flare}
\author{Pascal \surname{Saint-Hilaire}$^{1,2}$}
\author{Arnold O. \surname{Benz}$^1$}
\institute{ $^1$ Institute of Astronomy, ETH, CH-8092 Zurich, Switzerland \email{shilaire@astro.phys.ethz.ch}\\
$^2$ Paul Scherrer Institute, CH-5232 Villigen, Switzerland \\}
\runningtitle{Energy budget and imaging spectroscopy of a compact flare}
\runningauthor{Saint-Hilaire and Benz}

\begin{abstract}
We present the analysis of a compact flare that occurred on 2002/02/26 at 10:26 UT, seen by both RHESSI and TRACE.  
The size of the nearly circular hard X-ray source is determined to be 4.7($\pm$1.5)" from the modulation profiles of the RHESSI collimators. 
The power-law distribution of non-thermal photons is observed to extend down to 10 keV without flattening, and to soften with increasing distance from the flare kernel.
The former indicates that the energy of the precipitating flare electron population is larger than previously estimated: 
it amounts to $2.6(\pm0.8)\times 10^{30}$ erg above 10 keV, assuming thick-target emission. 
The thermal energy content of the soft X-ray source (isothermal temperature of 20.8($\pm0.9$) MK) and its radiated power were derived from the thermal emission at low energies. 
TRACE has observed a low-temperature ejection in the form of a constricted bubble, which is interpreted as a reconnection jet. Its initial energy of motion is estimated. Using data from both
satellites, an energy budget for this flare is derived. The kinetic energy of the jet bulk motion and the thermal and radiated energies of the flare kernel were more than an order of magnitude smaller than the derived electron beam energy.
A movie is available on the CD-ROM accompanying this volume.
\end{abstract}


\end{opening}

\pagebreak
\section{Introduction}
The energy of solar flares is commonly assumed to be magnetic in origin, but its release is still unclear. 
MHD theory of reconnection predicts equal shares of energy for local heating by electric resistivity and the motion of the plasma ejected from the reconnection region (e.g. Priest and Forbes, 2002). 
However, early observations of centimeter radio and hard X-rays (HXR) from flares made clear that a considerable fraction of the flare energy is initially transferred into energetic electrons (Neupert, 1968; Brown, 1971; Lin and Hudson, 1976) and possibly ions (Ramaty {\sl et al.}, 1995). 
These "non-thermal" particles carry a large fraction of the energy away from the flare region and deposit it partially in the chromosphere, where plasma is heated to tens of million degrees, rises into the corona and emits soft X-ray emission (Neupert, 1968). The knowledge of the energy content in the various forms of primary and secondary energy is essential in formulating flare scenarios and modeling flares.

The partitioning of the energy is not well known, as the exact evaluation of the various amounts of energy is hampered by observational limits. 
Early estimates by Strong {\sl et al.} (1984) of a simple flare using several instruments on SMM find a ratio of 1.3:1.7:1 for the distribution between electron beam, mass motion, and thermal energies, where the mass motions were measured in a Ca XIX line and may include also evaporative motions. 
It may be partially a secondary form of energy, derived from the kinetic energy of energetic particles. \inlinecite{deJager1989} have compared
beam and thermal energies of 19 flares with similar results.
Observations indicative of reconnection jets have been reported in the literature (e.g. Shibata {\sl et al.}, 1994; Pohjolainen {\sl et al.}, 2001; Zhang, Wang, and Liu, 2000). The identification as a reconnection jet in the corona was often based on the motion of dense material. 
As the process of reconnection is not predicted to substantially increase the density, these observations suggest that reconnection took place in a high-density region. 
This may not necessarily be the case in all flares. 
The heating of reconnection jets is not understood.
Although the plasma heated by resistivity is ejected, the jets also contain plasma at preflare temperature in the MHD scenario. Innes {\sl et al.} (1997) have reported reconnection jets in the quiet Sun having a temperature of a few $10^5$K. On the other hand, stochastic electron acceleration by transit-time-damping of low-frequency waves is currently the most preferred mechanism (Miller {\sl et al.}, 1997). 
It is expected to take place preferentially in the turbulent plasma of reconnection jets and may also heat them.

With the new generation of solar instruments in EUV lines, soft and hard X-rays, a more accurate determination of energies becomes feasible.  Of particular relevance is the Reuven Ramaty High Energy Solar Spectroscopic Imager (RHESSI) launched on February 5, 2002 (Lin {\sl et al.}, 2002). 
RHESSI's germanium detectors, flown in space for the first time, register photons in the energy range from 3 keV to 17 MeV with 1 keV resolution at low energies (Smith, 2002). For the first time it is possible to explore the low-energy limit of non-thermal electrons, where most of the beam's energy resides. 
Nine absorbing grids modulating by satellite rotation provide the basis for imaging. The new method allows reconstructing the image of a flare anywhere on the visible disc of the Sun with a resolution of about 2" at low energies (Hurford {\it et al.}, 2002). 
Thus, RHESSI's spatial resolution also allows determining the size of the high-temperature thermal flare plasma, necessary to estimate its energy content. 
Furthermore, RHESSI can model the thermal plasma by fitting the low-energy spectrum and determine its temperature and emission measure. 
The spatial resolution of an instrument like TRACE (Handy {\it et al.}, 1999) allows measuring the motion of coronal plasma at relatively low temperature predicted by MHD models for parts of the reconnection jets.

In this paper, we use the new capabilities for the first time to estimate the various flare energies in a well-observed, simple flare. The observations and some relevant analyses are presented in Sections 2 through 6, and in Section 7 the energies are evaluated and compared. A movie is available on the CD-ROM accompanying this volume.

\section{Lightcurves and other generalities}

\begin{figure}
\includegraphics[width=11cm]{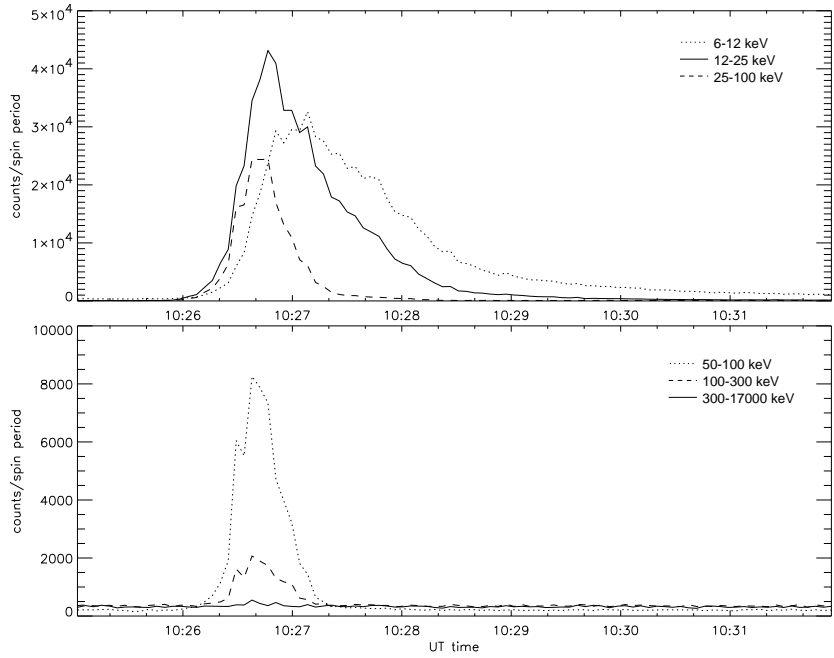} 
\caption{Lightcurves of RHESSI observations in uncalibrated counts per 4.35s rotation period. All RHESSI front detector segments were used.}
\label{lc1}
\end{figure}

A flare occurred near the western limb of the Sun on February 26,
2002, and was observed by both RHESSI and TRACE. NOAA/GOES reported it as a C9.7 
class flare.

\begin{figure}
\includegraphics[width=11cm]{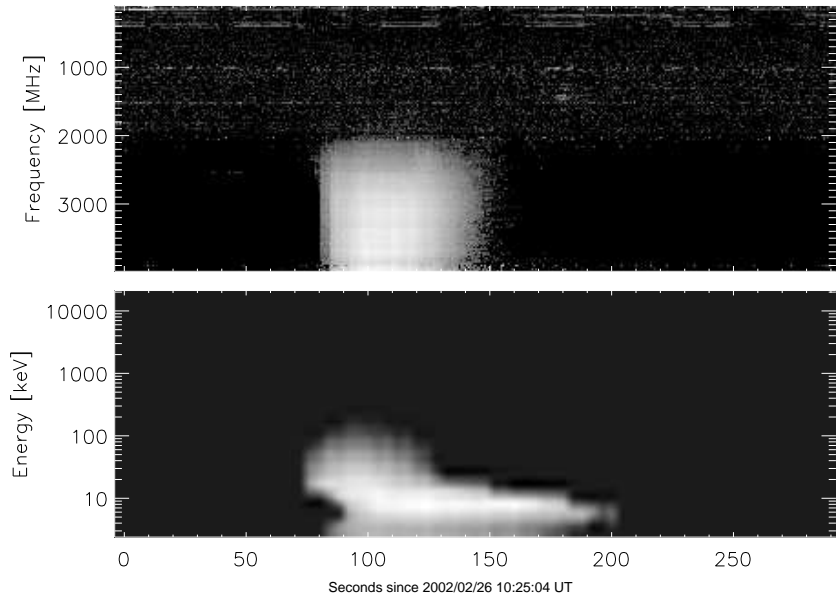}
\caption{Phoenix-2 radio (top) and RHESSI (bottom) spectrograms of the event.
Both are background-subtracted.} \label{spg}
\end{figure}

Figure \ref{lc1} shows that this event, while not the most powerful of flares, 
did have a non-negligible flux of high-energy photons (i.e. higher than 100 keV):
enough to allow images to be reconstructed (Fig. \ref{pte}). 
The Phoenix-2 radio spectrometer \cite{Messmer1999} saw gyrosynchroton radiation 
during the high-energy part of the HXR emission (Fig. \ref{spg}), but very little decimetric emission. It consisted of three 
reverse drifting type III bursts between 1.2-2.0 GHz and a regular type III burst (620-800 MHz) with possible
narrowband spikes at 850 MHz. All these faint coherent emissions (not visible in the overview of Fig. \ref{spg}) occurred after the HXR peak.

RHESSI's attenuator state was '1' (thin in, thick out) at all times. No decimation
occurred, RHESSI was 'in the clear' throughout the event (no SAA or eclipse). 
The flare took place when RHESSI was at high geomagnetic latitude (43 degrees).

As the average count rates per detector (total for all energy bands) was less than
6000 counts/s, no pile-up in the detector is expected \cite{Smith2002}. 
No data gaps were recorded, aside from the fact that detector 2 was turned off at the
time of the flare, and the presence of the usual {\it dropouts}. Dropouts are short data gaps ($\le$1s) that occur randomly in every RHESSI detector, and are most likely the result of cosmic ray hits that
momentarily saturate the detector electronics \cite{Smith2002} -- the imaging reconstruction software deals with those by ignoring them,
i.e. during those times, no weighted modulation pattern contribution is added to the  back-projected map.

\section{Source size}

\begin{figure}
\includegraphics[width=11cm]{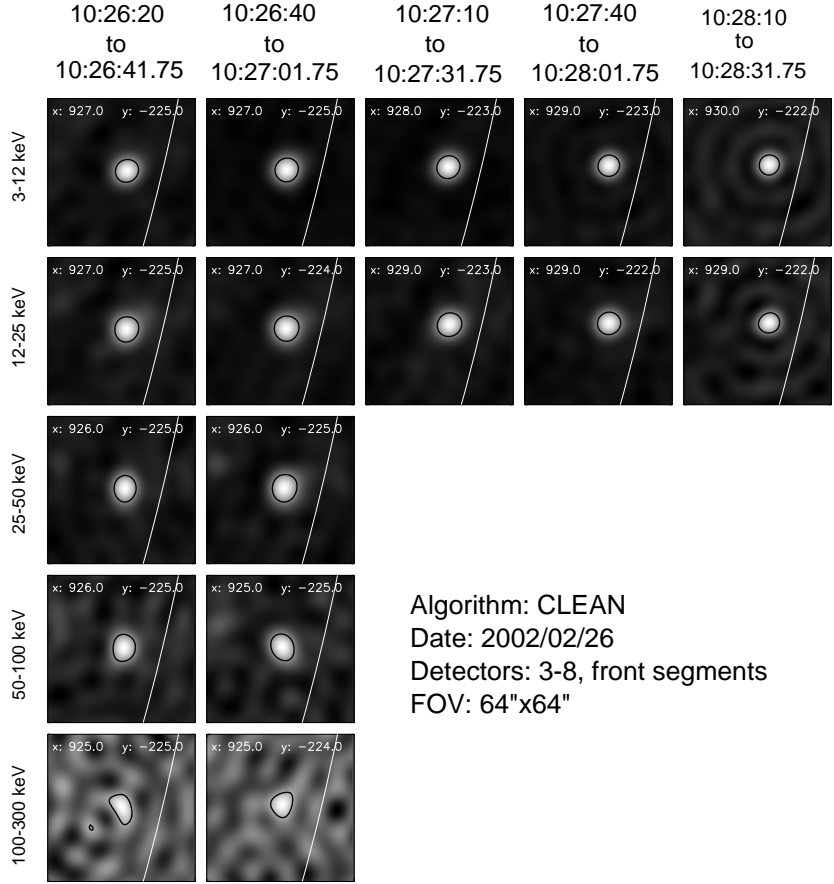}
\vskip5mm
\caption{Panel of images: time (horizontal) vs. different energy bands (vertical).
The black contour show the 50\% level of the image.
The white line is the photospheric limb of the Sun.
The white numbers on each panel refer to the position, in arcseconds from Sun center,
of the brightest pixel in the image.
} \label{pte}
\end{figure}

Figure \ref{pte} shows the flare at different times and energies. 
As the flux diminishes with higher energy, the brightness of the images has been adjusted. 
It can be seen that its spatial shape is nearly circular and remains practically constant during the HXR emission
(special care has been taken to ensure that the {\it same} aspect solution was used for all the images.).
The deviation from circular at 100 -- 300 keV is not to be considered statistically significant. 

\begin{figure}
\rotatebox{90}{
\includegraphics[width=11cm]{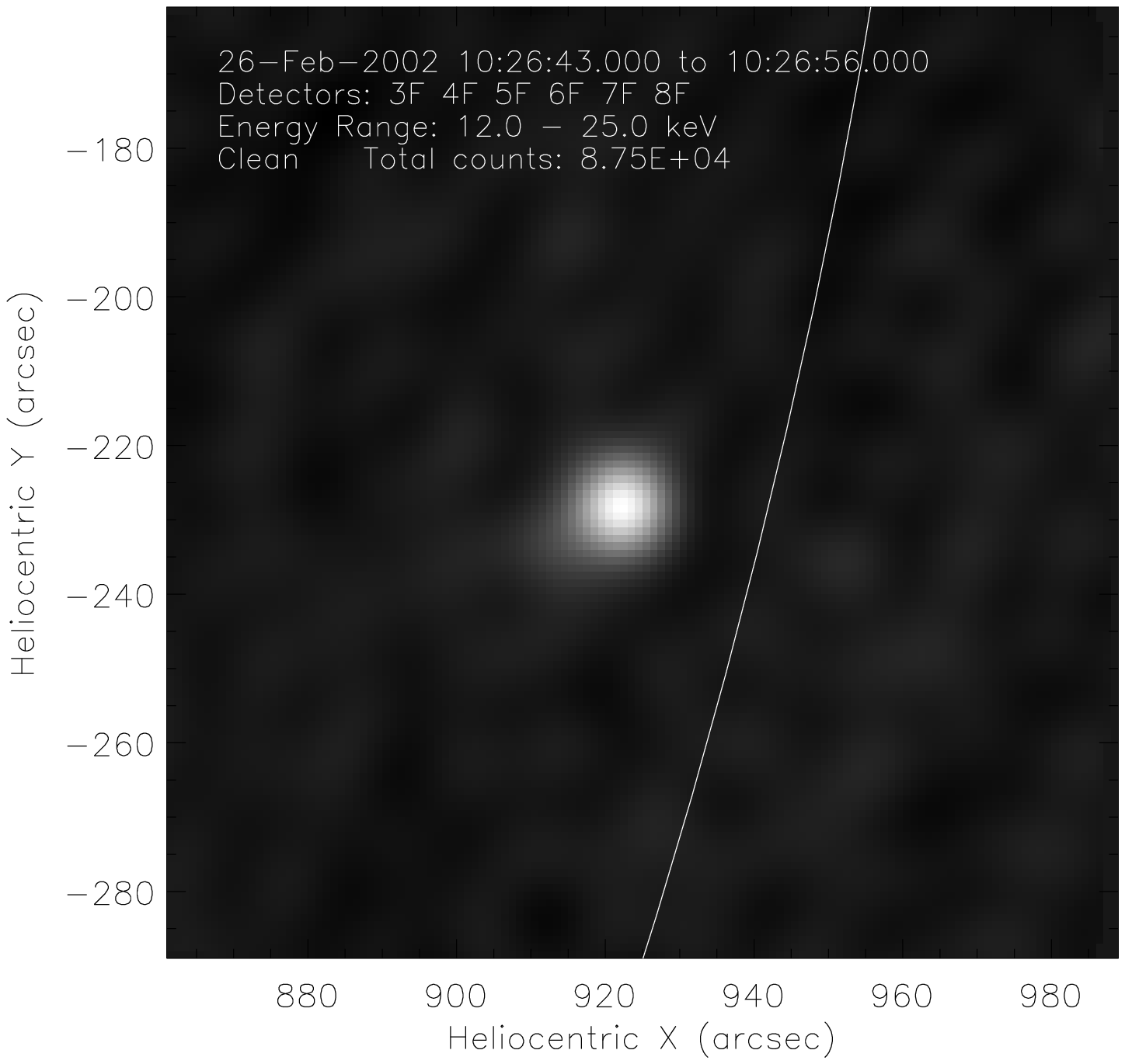}}
\caption{RHESSI CLEANed image made at peak 12-25 keV flux time (10:26:43 to 10:26:56 UT).}
\label{peaktime}
\end{figure}

\begin{figure}
\includegraphics[width=11cm]{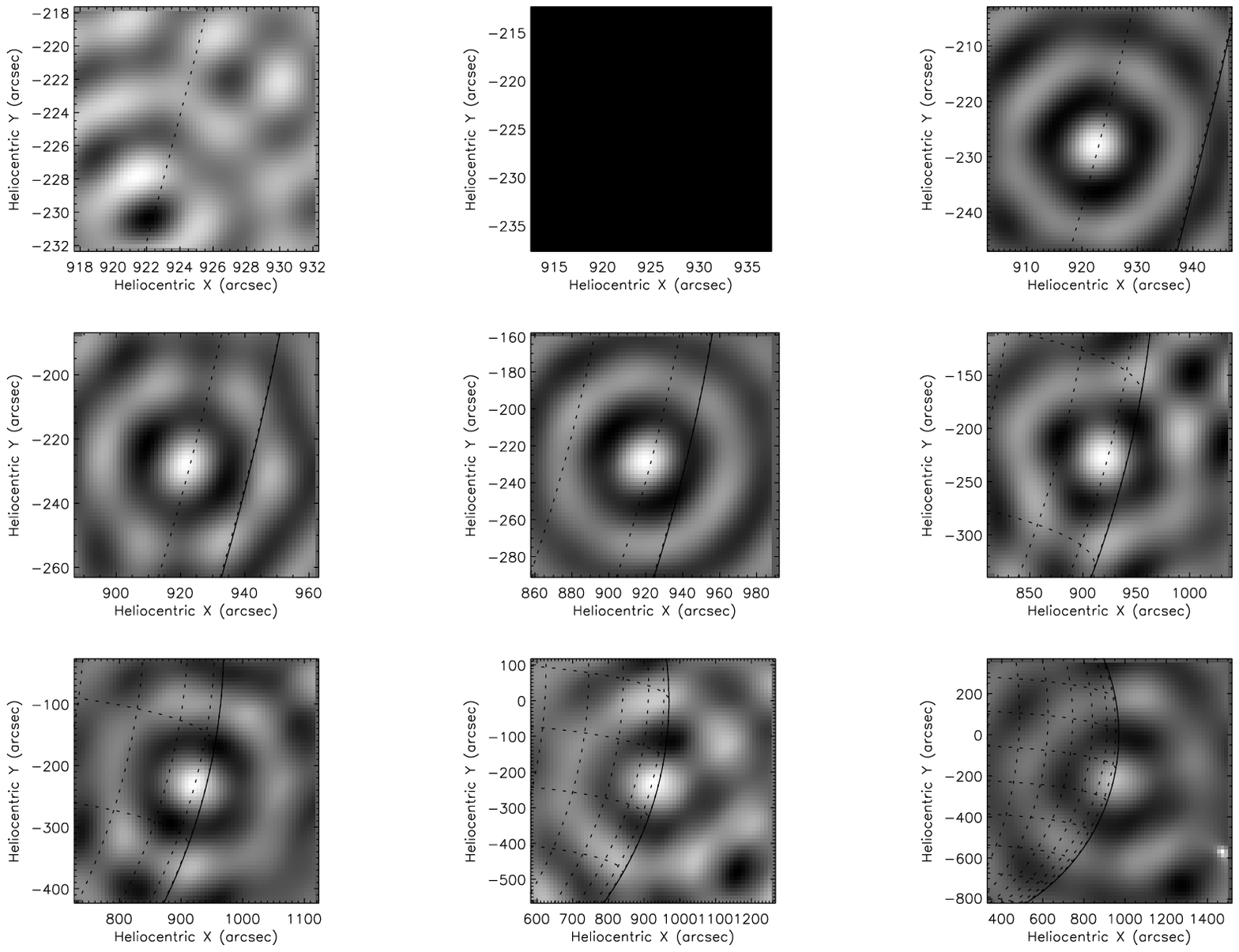}
\caption{Back-projected images for each subcollimator (1 to 9), other parameters are the same as for Figure \ref{peaktime}. The field of view and the pixel size are adjusted to the resolution of each subcollimator, such that the spatial sampling is twice the resolution.}
\label{sc_short}
\end{figure}

Figure \ref{peaktime} is a close up of the flare, as seen by RHESSI. It is the result of careful elimination of unbeneficial collimators and detectors. 
The first panel of Figure \ref{sc_short} shows that subcollimator 1 would not have contributed in a useful manner to the overall image: a
pattern of minima and maxima of size and spacing comparable to that subcollimators's FWHM. Taking a longer accumulation time, i.e.
the whole time interval when HXR counts were above background and the source spatially stable (10:26:20 to 10:27:10 UT), does
not yield a better result, even though the estimated total number of counts above background ($\sim$25000) should have been sufficient. 
We conclude that subcollimator 1 has over-resolved the source. Detector 2 was unfortunately turned off at that time.  
Detector 9 (FWHM of 186") was also removed, as its low resolution does not add any new features to the image. Hence, our imaging capabilities
are limited by subcollimator 3, which has a FWHM point spread function of $\sim$6.9". 

Table \ref{SourceSize} lists the results of several methods used to determine the source size. The `modulation amplitude' method will
be described in this section, whereas the `imaging spectroscopy' method will be described in section 4.
A source imaged with RHESSI appears larger than its true size $\sigma_s$, the standard deviation of an equivalent gaussian. 
The observed source size $\sigma_{\rm obs}$ results from convolution with the point spread function $\sigma_{\rm psf}$ of the instrument, 
where $\sigma_{\rm s} = \sqrt{\sigma_{\rm obs}^2 - \sigma_{\rm psf}^2}$,
assuming that both source and point spread function have gaussian shapes. 

The FWHM of the source was determined from the 50\% contour of its image reconstructed by back projection, CLEAN and Maximum Entropy Method (standard RHESSI software)\footnote{http://hesperia.gsfc.nasa.gov/rhessidatacenter/}. 
The FWHM of the point spread function of each RHESSI collimator was measured the same way. 
For all collimators, the two values were found to be the same within the errors. 
Thus only an upper limit of the true source can be determined. 
The low value of subcollimator 3 is surprising, but within the statistical error.
The very small FWHM obtained by MEM-Sato may be caused by what has been dubbed `super-resolution', and is not to be trusted. 

\begin{table}
\begin{tabular}{l|c|c|}\hline 
Method 					& convolved FWHM	& deconvolved or 	\\
					&			& true FWHM		\\
\hline
FWHM of maps:				&			&			\\
`back proj.'/CLEAN using SC 3 		& 6.5$\pm$0.5" 	& $\lapprox$3.2"	\\
`back proj.' using SC 4 			& 11.0$\pm$1.0" 	& $<$5.7"	\\
`back proj.' using SC 5 			& 19.8$\pm$1.8" 	& $<$7.5"	\\
MEM Sato, using SC 3-8			& 2.9$\pm$0.2"		& -			\\
MEM Sato, using SC 3			& 7.7$\pm$0.2"		& 3.4"	 		\\
\hline
imaging spectroscopy & 11$\pm0.5$" 			& -		\\
(CLEAN using SC 3-8) &					&		\\
\hline
modulation amplitude 	&  		 	& 4.7$\pm$1.5"				\\
\hline \end{tabular}
\caption {Size of non-thermal HXR source (12--25 keV), derived by different methods. 
The error of the convolved size from RHESSI maps indicate the total range.} 
\label{SourceSize}
\end{table}

It is clear from the different imaging algorithm that the source size must be smaller
than subcollimator 3's FWHM of $\sim$6.9", but larger than subcollimator 1's FWHM of $\sim$2.3".
This yields a range of possible values for the source's FWHM of 2.3"--6.9", or 4.6($\pm$2.3)".
To try to determine the size more accurately, a closer look at modulation profiles was taken.

\begin{figure}
\tabcapfont
\centerline{\includegraphics[width=12cm]{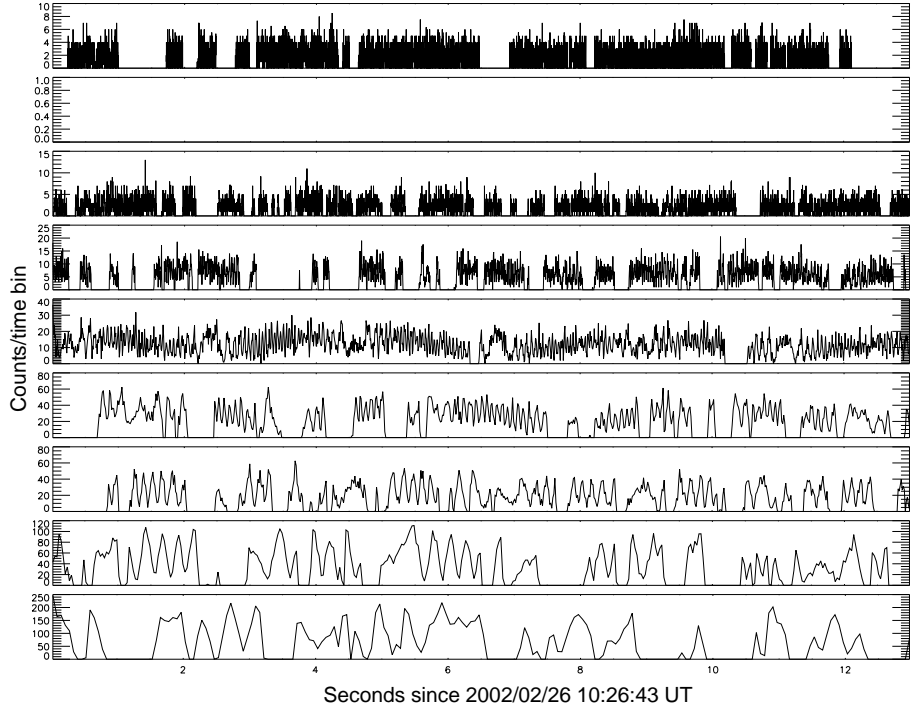}}
\caption{Sample of modulation profiles for each subcollimator (1 through 9: from top to bottom) 
for the time interval used for the image in Figure \ref{peaktime}.}
\label{modprof}
\end{figure}

Figure \ref{modprof} displays the modulation profiles that were used to produce Figure \ref{peaktime}. 
From those, the {\it relative modulation amplitudes} \cite{es_gh2002} were determined.
The relative modulation amplitude $A$ is defined as follows:

\begin{equation}
	A = \frac{1}{M} \frac{C_{max}-C_{min}}{C_{max}+C_{min}} =  \frac{1}{M} \frac{C_{max}-<C>}{<C>}
	\label{eq_A}
\end{equation}
$C_{max}$ and $C_{min}$ are the maximum and minimum counts per time bin in a modulation cycle.
$M$ is the {\it maximum modulation amplitude}. It is determined by the angle of incidence on the grid as well as the 
effective slit/pitch ratio. The software calculates it for each time bin.
\inlinecite{es_gh2002} give an analytical formula for $A$ in the case of gaussian sources:

\begin{equation}
	A = e^{-0.89 \left( \frac{\Delta\theta_s}{\Delta\theta_{coll}} \right)^2},
	\label{eq_th_A}
\end{equation}
where $\Delta\theta_s$ is the source's FWHM, and $\Delta\theta_{coll}$ is the collimator's FWHM. 
Thus, the modulation disappears gradually as the source dimension exceeds the collimator resolution. 

\begin{figure}
\tabcapfont
\centerline{%
\includegraphics[width=11cm]{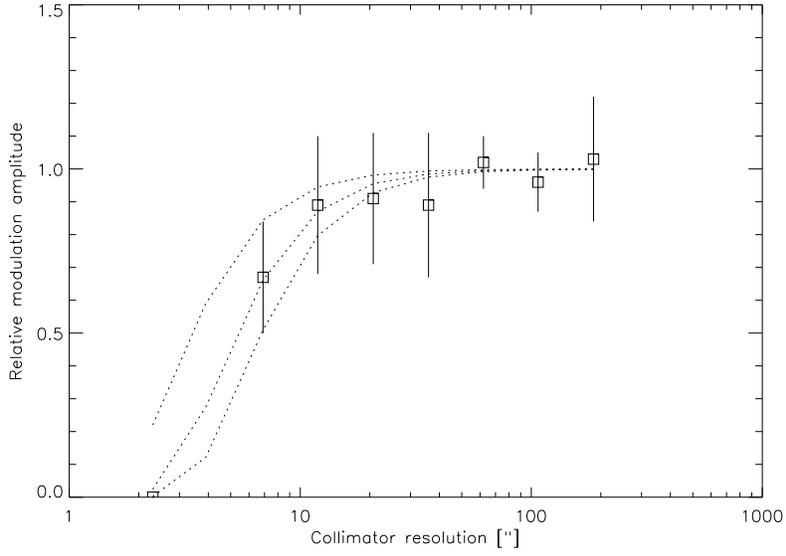}
}
\caption{Observed relative modulation amplitude vs. subcollimator resolution, from the modulation profiles of Fig. \ref{modprof}.
The dotted curves were computed using Eq. (\ref{eq_th_A}). The upper one for a source size of 3" FWHM, the middle one
for a size of 4.7" FWHM, and the lower one for a size of 6" FWHM.}
\label{ModFlux}
\end{figure}

\begin{figure}
\tabcapfont
\centerline{%
\includegraphics[width=11cm]{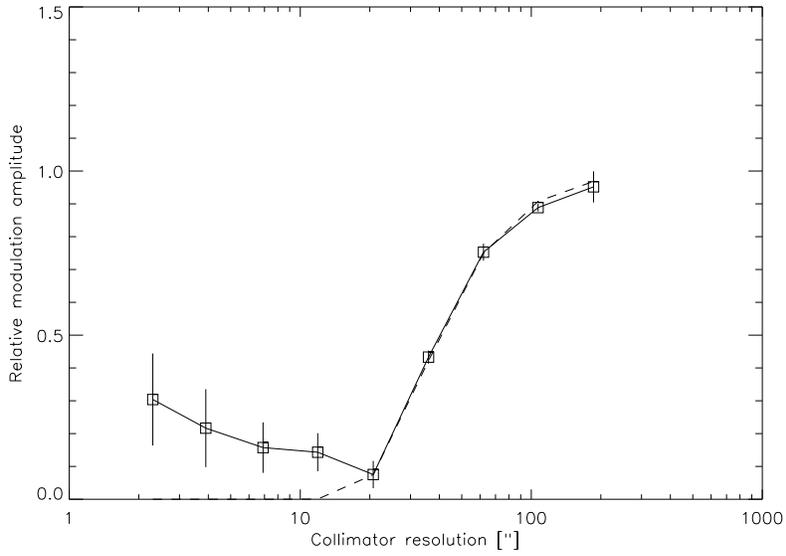}
}
\caption{Relative modulation amplitude vs. collimator resolution (FWHM), for a simulated 2-D circular gaussian source of size $\sigma$=15" (= 35.3" FWHM).
The solid line links the data points, whereas the dashed line was computed using Eq. (\ref{eq_th_A}).}
\label{SimModFlux}
\end{figure}

$C_{max}$ and $C_{min}$ were determined manually and averaged over several modulation cycles (from 3 for subcollimator 9 to 60 for subcollimator 3, also yielding a standard deviation) in the modulation profiles (Fig. \ref{modprof}).
Thus, the contaminating effects of dropouts have been removed. The results are displayed in Figure \ref{ModFlux}.
The relative modulation amplitude was set to zero for the first (finest) subcollimator, where photon fluctuation was clearly dominant.
The relative modulation amplitude of subcollimator 2 was ignored, as it is not available.
Few modulation cycles without dropouts were available (Fig. \ref{modprof}) for subcollimator 9, resulting in a large standard deviation.
Otherwise, the error bars increase with decreasing subcollimator coarseness. This is because the size of the time bins that were used also decreased
with decreasing subcollimator coarseness, thus increasing the effects of photon counting noise.
Comparing the data points with the theoretical curves, a source size of 4.7$\pm$(1.5)" is assumed, and will be used in the numerical computations of section 7.

The method was tested on gaussian sources of different sizes using the RHESSI simulation software tools. The match
is almost perfect for regions with low photon fluctuations. Figure \ref{SimModFlux} is one such plot, made with $5 \times 10^{5}$
$photons/s/detector$, for a source size of 35.3" FWHM, and in the same 12-25 keV energy band as used previously. The existence of
non-zero relative modulation amplitudes at low collimator resolution is due to photon counting noise, and the manual technique
for finding the peaks, which does not make any use of the phase (as a forward fitting method would).

The 4.7($\pm$1.5)" source size derived from the modulation amplitude method is consistent with the upper limits derived
using other methods (Table \ref{SourceSize}), and is only marginally better than the 4.6($\pm$2.3)" range of possible values previously
derived, because of the low count rate, which makes photon counting noise significant.

\section{RHESSI imaging spectroscopy}

Imaging spectroscopy is limited by photon-counting noise. 
Hence, we will simply concentrate on doing imaging spectroscopy during the peak HXR flux of the flare. As the images obtained are not fully calibrated through the spectral response matrix \cite{Schwartz2002}, only energies
above 10 keV and below 100 keV were considered.

\begin{figure}
\tabcapfont
\centerline{%
\includegraphics[width=11cm]{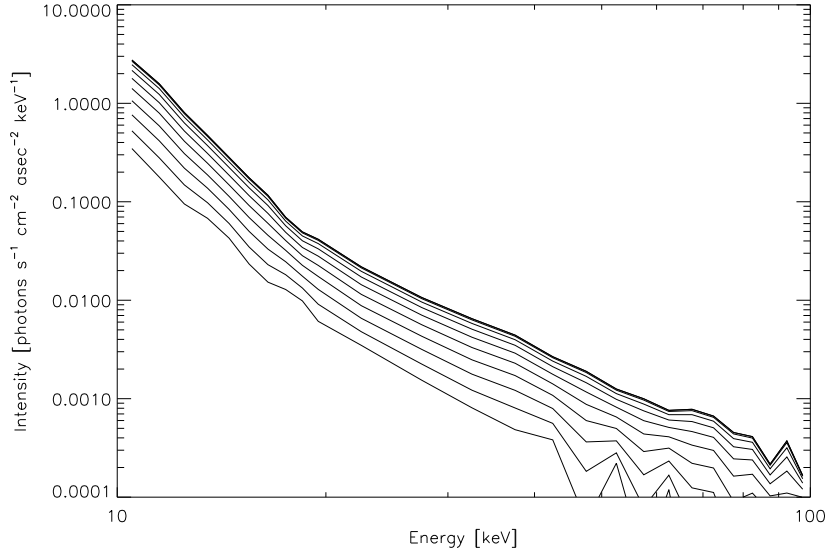} 
}
\caption{Imaging spectroscopy I: spectra at different distances from flare center position at the time of peak HXR flux.
Spectra from top to bottom were taken at increasing distances, 1" increments, starting at 0" for the topmost spectrum.
At high energies and large distances, photon fluctuation effects become important. 
}
\label{ImgSpec}
\end{figure}

\begin{figure}
\tabcapfont
\centerline{%
\includegraphics[width=11cm]{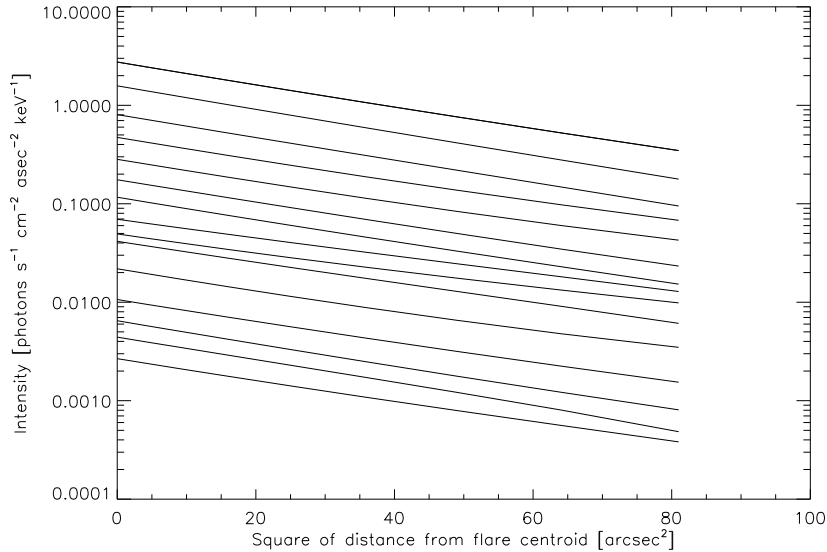} 
}
\caption{Imaging spectroscopy II: flux vs. distance from flare center position, for different energy bands. 
The flux in different energy bands (from 10 to 45 keV) are ordered with increasing energy, from top to bottom.
From 10 to 20 keV, 1-keV bands were used. From 20 to 45 keV, 5-keV bands were used.}
\label{ImgSpec2}
\end{figure}

The flare was imaged (using CLEAN) between 10:26:43 and 10:26:56 UT, from 10 to 100 keV (using 1-keV energy bins from 10 to 20 keV, then 5-keV bands), and
using detectors 3, 4, 5, 6, and 8. The image was over-sampled by taking 1" pixels. The 26 images thus obtained are not shown here.
A `crosshair' of pixels (10 vertically, 10 horizontally), centered on the flare, was considered. 
Figures \ref{ImgSpec} and  \ref{ImgSpec2} show results obtained by averaging pixel fluxes at equal distance from the flare's center.

Figure \ref{ImgSpec} shows that the portion of the spectrum above the thermal bremsstrahlung component is close to a power-law, with photon spectral indices
varying from $3.4 \pm 0.2$ (flare center, associated with the brightest pixel in the map) to $3.9 \pm 0.2$ (at 10" from flare center), with the fits made in the 20-40 keV energy bands.

The hardening of the spectral index towards the center of the flare kernel may be due to a harder electron distribution (more fast electrons) 
near the center of the flare. 

This observation does not support proposed models such as a superposition of thermal distributions (Brown, 1974)
mimicking a power-law of energy spectral index, e.g. $\gamma = \frac{1}{2} + \frac{3}{\eta}$ from a sum of thermally emitting spherical co-centric shells with temperatures $T(r) = T_0(\frac{r_0}{r})^\eta$.

The flare had a gaussian shape at all energies (Fig. \ref{ImgSpec2}). The slope (=-$1/2 \sigma^2$)
was about $-0.023 \pm0.002$, implying an apparent source sigma of $4.7(\pm0.2)"$, or FWHM of $11(\pm0.5)"$. 
Note that the apparent source size is again a convolution of the true source size with the point spread function of the imaging instrument. 

\section{Spectral features}

\begin{figure}
\includegraphics[width=11cm]{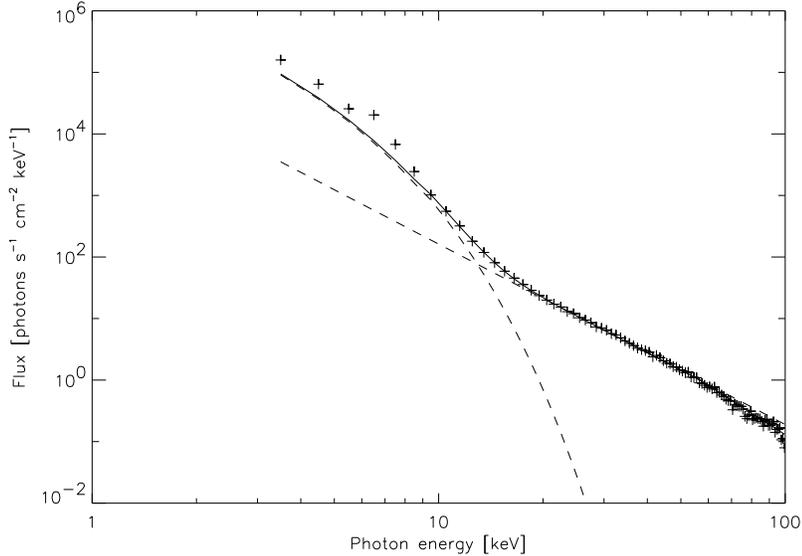}
\caption{Spectrum accumulated from 10:26:05 to 10:27:23 UT, during enhanced HXR $>$25 keV flux.
Only the front segment of detector 4 was used. The `+' symbols are the data points. 
A fit with a thermal bremsstrahlung and a power-law is also drawn (solid line).
} 
\label{peakSPEX}
\end{figure}

Figure \ref{peakSPEX} shows a spatially integrated spectrum accumulated during the peak of the HXR flux. As RHESSI's spectral response
below 10 keV is not yet completely known (particularly when a shutter is in, as is the case here), spectral fitting (using the full spectral response matrix) has been done on energies above 10 keV.
The data points were fitted using the SPEX\footnote{http://hesperia.gsfc.nasa.gov/rhessidatacenter/spectroscopy.html} software, with a thermal free-free bremsstrahlung and double broken power-law model.

The thermal bremsstrahlung component of the fit model yields a temperature of $T=19.7 \pm 1.0$ MK, and an emission measure of $EM\sim 0.2 \times 10^{49} $cm$^{-3}$
(accurate to within a factor 2), assuming an isothermal source. 
The temperature derived from GOES-8 3-second data peaks at 10:26:51 UT at 16.7 MK, with an emission
measure of $0.6 \times 10^{49} $cm$^{-3}$.

The power-law component had a photon flux at 50 keV of $1.5 \pm 0.2$ photons s$^{-1}$cm$^{-2}$keV$^{-1}$, 
and a spectral index of $3.0 \pm 0.1$. This value is not significantly different than the ones derived in the previous
section, where a different time interval was used, and where only energies in the 20-40 keV range were considered.
A break in the power-law is located at $54 \pm 3$ keV.
After this break, the photon spectral index is $3.5 \pm 0.1$. 
Breaks are commonly observed (see, e.g. \opencite{LinSchwartz1987}) and do not significantly influence the
energy budget (section 7).


An emission volume V can be inferred from the size of the source observed with imaging: 
$V=\frac{4 \pi}{3} R^3$, where R is the size of the source. With $R=2.35(\pm 0.75)$", one finds an emitting volume of $V=2.1 \times 10^{25} $cm$^{-3}$,
with $0.8-4.8 \times 10^{25} $cm$^{-3}$ the range of possible values.

The thermal energy and density can then be calculated, using:

\begin{equation}	
	E_{th} = 3\cdot k_BT \sqrt{EM\cdot V \cdot q}\ \ \ ,
\label{eth}
\end{equation}
\begin{equation}	
	n_e = \sqrt{\frac{EM}{V} \cdot q}\ \ \ ,
\label{EQ_ne}
\end{equation}
where $EM$, the emission measure, and $T$, the temperature, are the ones derived above. The filling factor is represented by a fraction q.
An inhomogeneous medium possesses less thermal energy than a homogeneous one, for the same temperature and emission measure. 
$q=1$ is assumed throughout this paper.
Eq. (\ref{EQ_ne}) yields a density of $n_e = 3.3 \times 10^{11} $cm$^{-3}$ (range: $1.5-8.1 \times 10^{11} $cm$^{-3}$), comparable
to what is derived from TRACE in section 7.

For determination of the thermal energy in the flare (section 7), temperature and emission measure determined at 
the peak of the soft X-ray (SXR, $<$12 keV) flux are needed. Spectral fitting done around 10:27:10 UT, accumulated over three
RHESSI rotations of 4.35s, yields $T$=20.8$\pm(0.9)$ MK and $EM$=2.9$(\pm0.5) \times 10^{48} $cm$^{-3}$.

\section{TRACE images with RHESSI overlays}

\begin{figure}
\tabcapfont
\centerline{%
\begin{tabular}{cc}
\includegraphics[width=5cm]{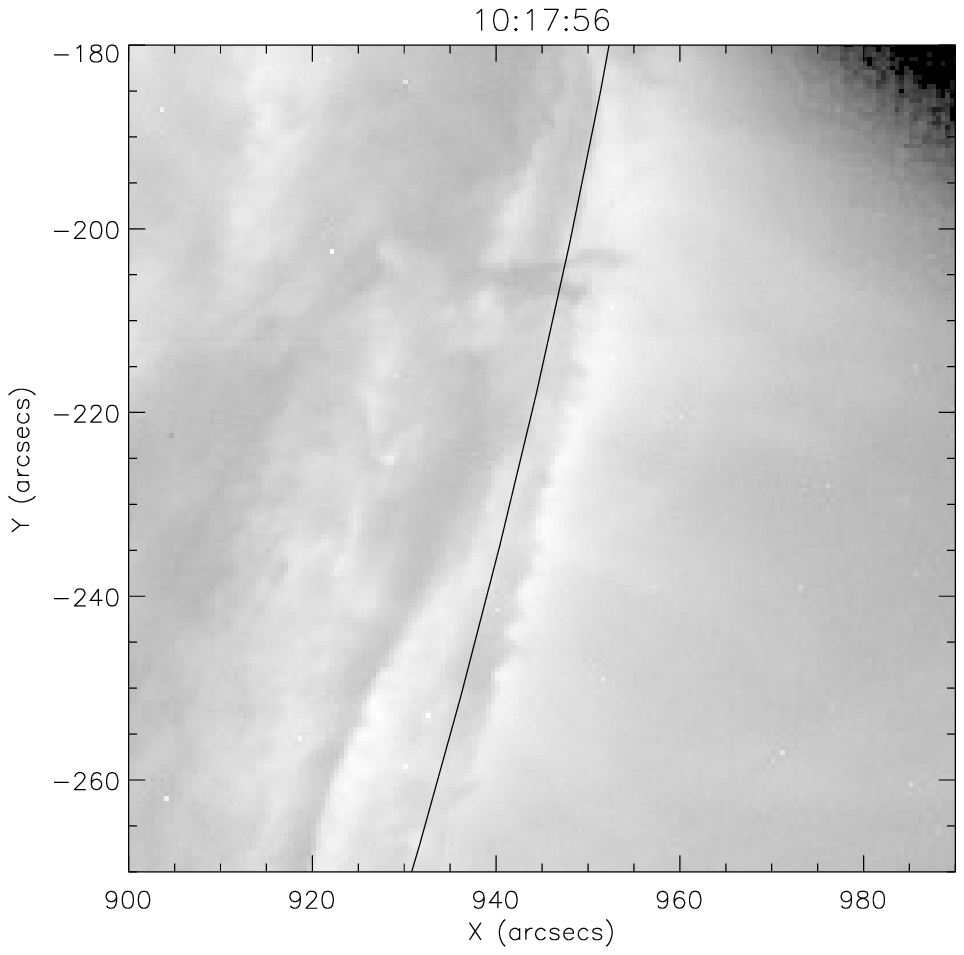} &
\includegraphics[width=5cm]{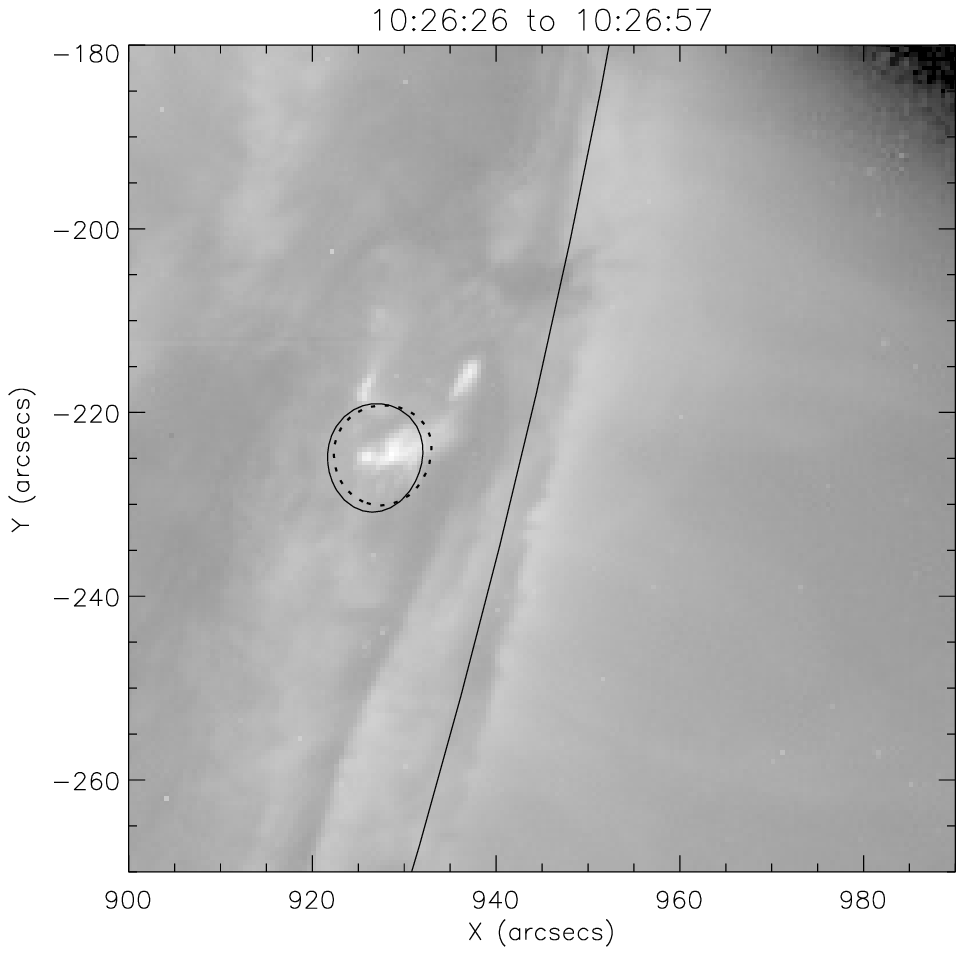} \\
\includegraphics[width=5cm]{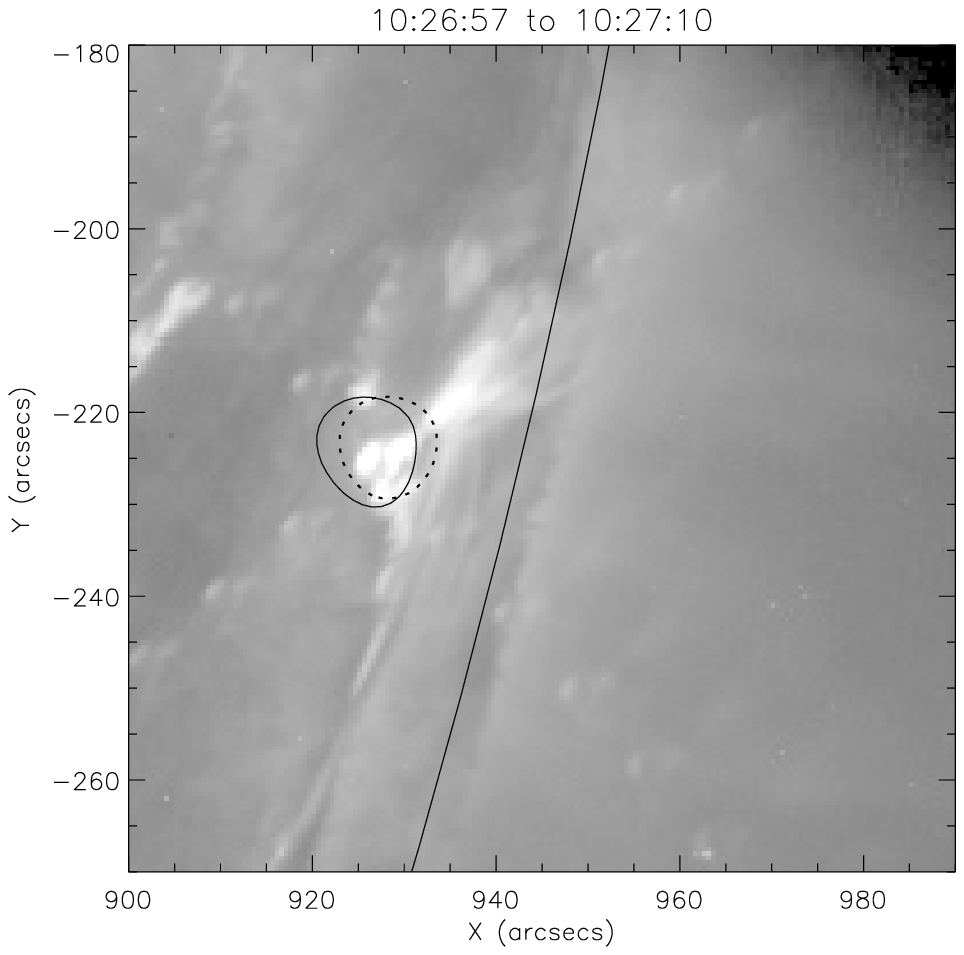} &
\includegraphics[width=5cm]{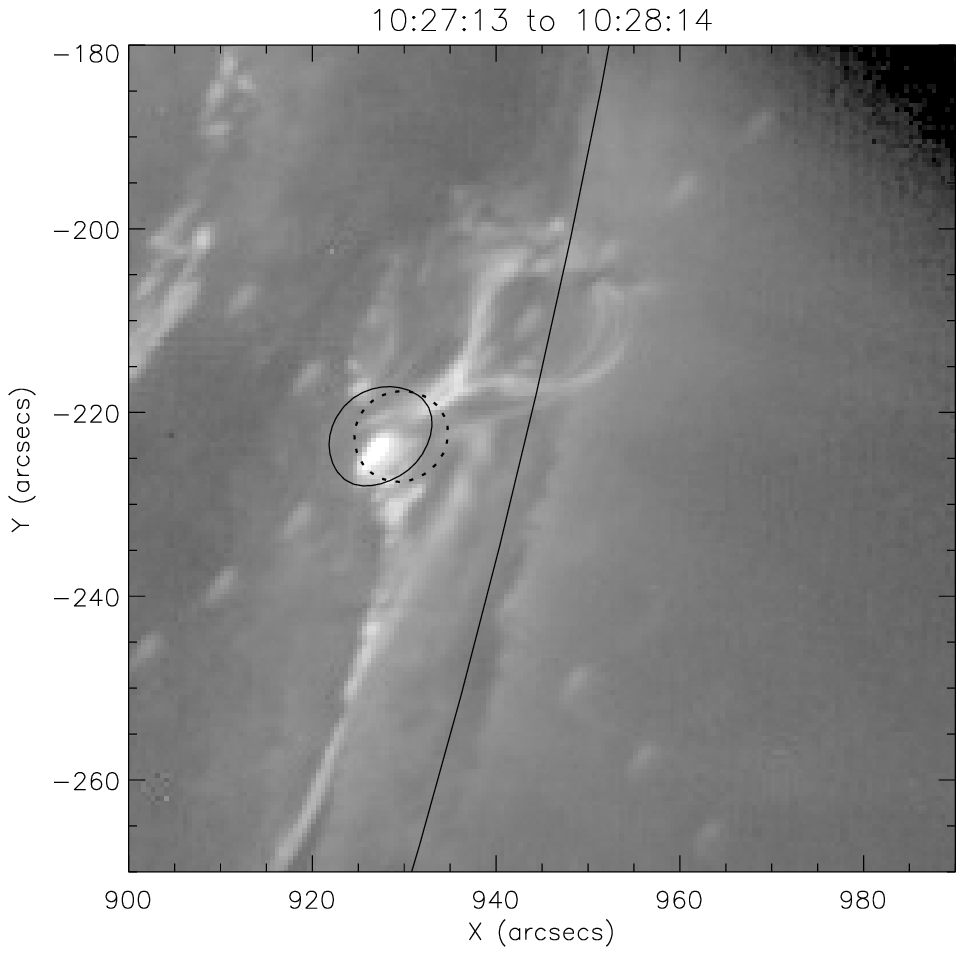} \\
\includegraphics[width=5cm]{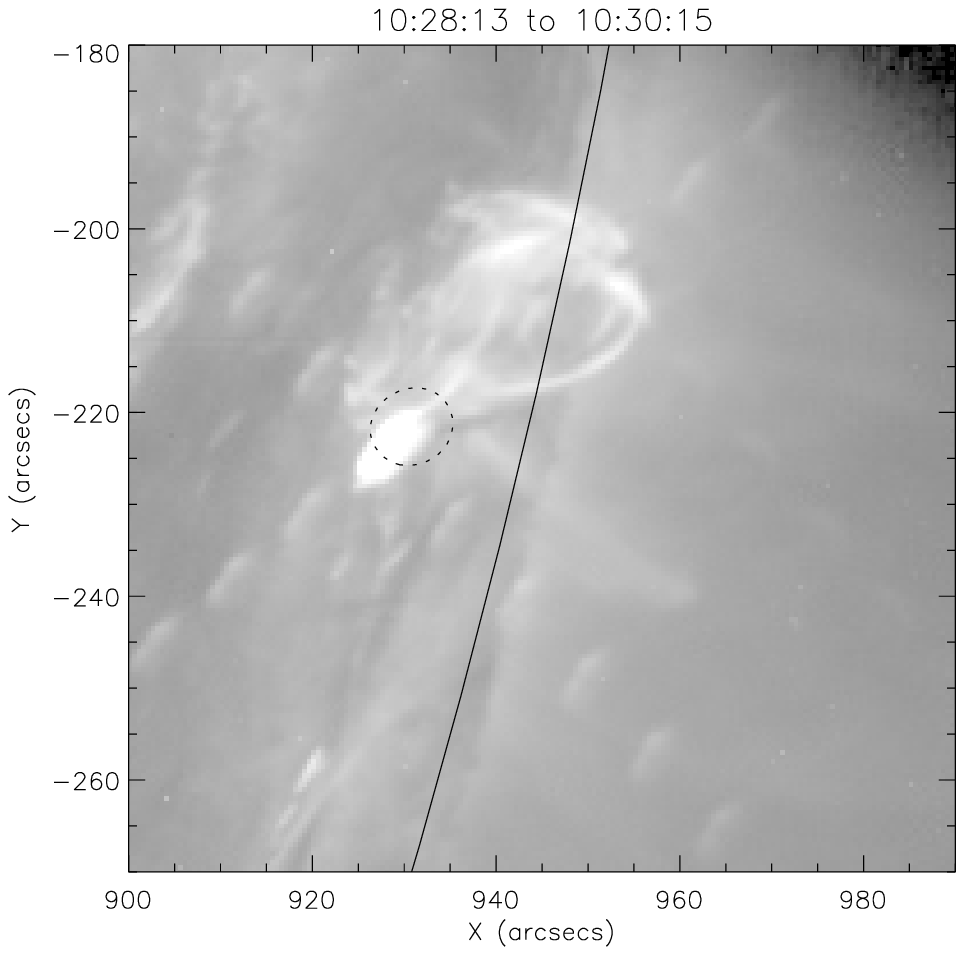} &
\includegraphics[width=5cm]{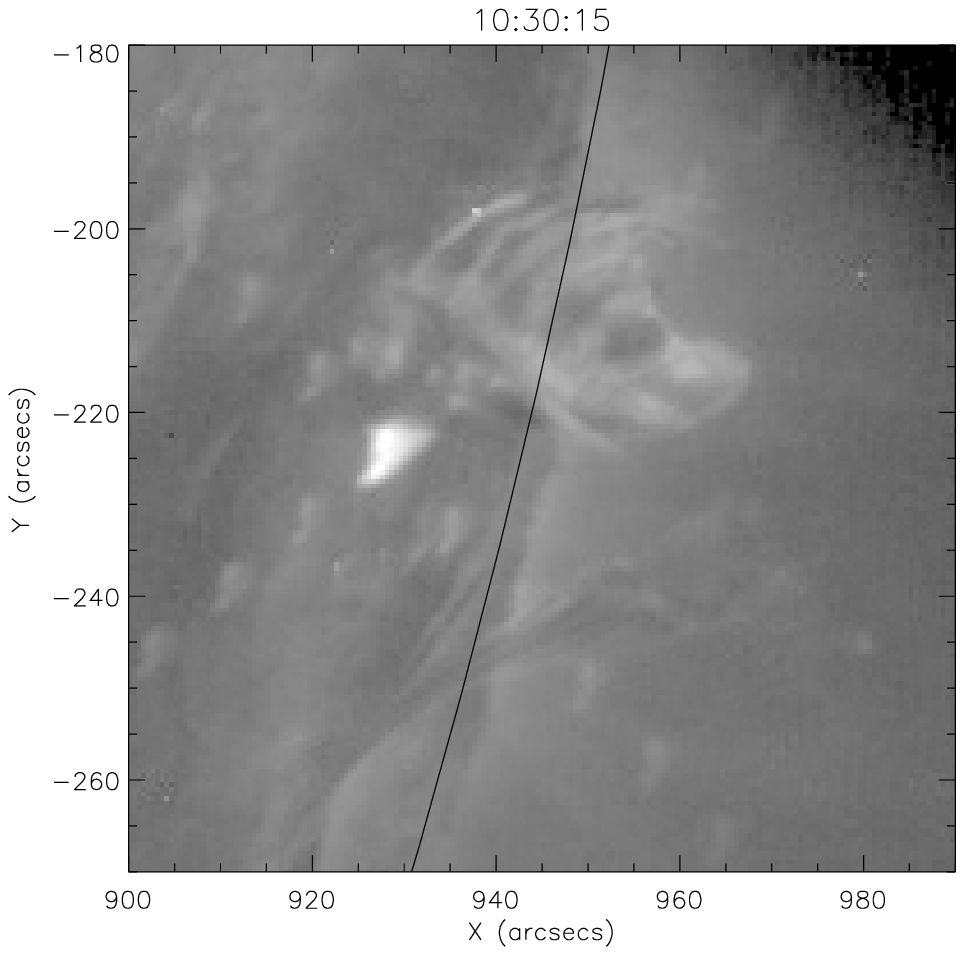} \\
\end{tabular}}
\caption{TRACE images at 195\AA\ with RHESSI overlays of different energy bands. The contours correspond to the 50\% level. 
The dotted black contour corresponds to the 12-25 keV band, the full black contour to the 25-50 keV band.
As shown in Fig. \ref{pte}, the 3-12 and 12-25 keV images differ by less than 1" throughout the flare, as do
the 25-50, 50-100, and 100-300 keV images.
The first image shows the region of interest before the flare. The second one was taken during the rise of the HXRs. 
The third one is between HXR and SXR peaks,
the fourth one between SXR (3-12 keV) and EUV peak (flash phase). The last two were taken during the decay phase.}
\label{TRACEimgs}
\end{figure}

Figure \ref{TRACEimgs} shows TRACE images with RHESSI overlays at different energies. 
The TRACE images had exposure durations of 20--30 seconds, except for the fourth one (EUV peak), which had an exposure duration of 8 seconds. 
All TRACE images have been translated 10" northerly, to align with RHESSI.
The EUV band pass is dominated by a spectral line of Fe XII (195 {\AA}) having a maximum emissivity at 1.4 MK. 
At high temperatures (15-20 MK), an Fe XXIV line (192 {\AA}) appears and may add significant flux (the filter's response around 15-20 MK
is still two orders of magnitude less than at 1.4 MK). 

The RHESSI images' accumulation times (as labelled on top of each image) correspond loosely to the time difference
between TRACE images (in integer multiples of the spin period of 4.35 s), and were all made using the CLEAN reconstruction
algorithm, and subcollimators 3 to 8. 
Again, care has been taken to use the same aspect solution for all RHESSI images.

\begin{figure}
\centerline{
\includegraphics[width=5cm]{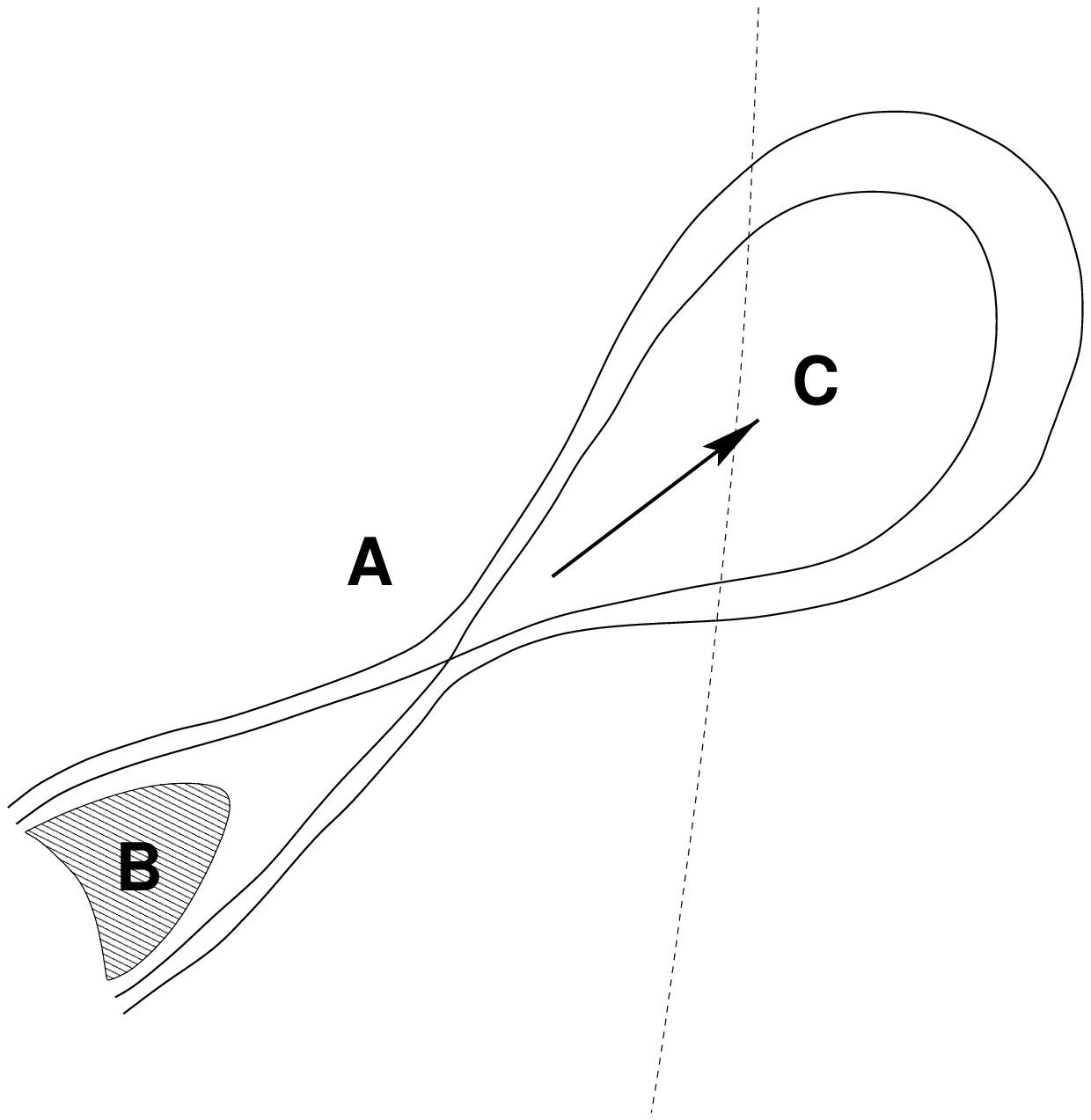} 
}
\caption{Schematic drawing of field lines and interpretation of TRACE image:\ \ \ \ \  A = X-point (reconnection site), B = flare kernel, C = upward reconnection jet.}
\label{scenario}
\end{figure}

The TRACE observations clearly show an ejection occurring with the flare, starting in the second image of Fig. \ref{TRACEimgs}. 
Later it develops into the shape of a bubble, which gets constricted at the bottom (best visible in Fig. \ref{TRACEimgs}, fourth image). 
It does not rise beyond the TRACE field of view, but becomes turbulent (see movie available on the CD-ROM).
The formation, development and constriction of the bubble is suggestive of a reconnection jet scenario speeding from the apparent X-point located at (930,-220) arcseconds from Sun center in Fig. \ref{TRACEimgs} (fourth image). 
The proposed scenario is depicted in Fig.  \ref{scenario}. 

RHESSI overlays show that the HXR (12-25 and 25-50 keV) are emitted in the flare kernel (marked by B in Fig.  \ref{scenario}). 
At the peak flux of HXR $>$25 keV, the peak positions of the sources $>$25 keV are shifted to smaller radial distances from Sun center ($\sim$ 1-2", best seen with Figure \ref{pte}), consistent with 
the interpretation that, after being accelerated by a reconnection event, the mildly relativistic electrons precipitate in the lower corona by emitting thick-target HXR radiation. 
The electrons with the higher energies will lose their energy (mostly via Coulomb collisions) only in the deeper, denser chromosphere.
The hot plasma being heated by the precipitation of electrons ($<$25 keV overlays or images in Figure \ref{TRACEimgs} and \ref{pte}) was apparently in the same volume.
After 10:27:20 UT, when the HXR emission above 25 keV ceases, a slow outward (NW direction) movement of the 3--12 and 12--25 keV (thermal) sources is witnessed,
at a speed of about 30 km/s.


\section{Energy budget}

Table \ref{Ebudget} summarizes the different energies found. In this Section, their derivation from the data is described.

\emph{Kinetic energy of precipitating electrons $E_{\rm kin,beam}$:} Assuming that the power-law component in our spectra is related to the HXRs emitted by precipitating
electrons in the lower corona (thick target), an electron distribution can be determined from the relations in Brown (1971) and Hudson (1978):

\begin{equation}
\frac{\partial{^2 N_e}}{\partial{E} \partial{t}} = 3.28 \times  10^{33} \, b(\gamma) \, \frac{F_{50}}{50^{-\gamma}} E^{-\gamma-1} \ \ \  [e^{-} \, s^{-1} \, keV^{-1}]
\label{HudsonFormula}
\end{equation}
where $\gamma$ is the photon spectral index, $F_{50}$ is the photon flux (in photons s$^{-1}$cm$^{-2}$keV$^{-1}$) at 50 keV,
E is the electron kinetic energy in keV, and b($\gamma$) is equal to 7.05 for $\gamma$ = 3.


At low electron energies, the distribution must become flatter, reducing the photon distribution below the cut-off energy.
Power-law distributions with low energy cut-offs have been fitted to the spectrum (Fig. \ref{peakSPEX}).
A cut-off above 10 keV decreases the match with observations, below 10 keV it has little influence on the fit of the photon spectrum.
In the absence of effects that enhance the low-energy photon flux (such as non-uniform target ionization, 
\opencite{Kontar2002}), we conclude that the low-energy cut-off for the photon power-law spectrum is $\leq$ 10 keV.
In the following, an electron power-law cut-off energy of 10$\pm1$ keV is assumed.

Determining the number of electrons $\ge$10 keV that precipitated, as well as their total kinetic energy, is simply
done by integrating over time and energy (from 10$\pm1$ keV to $\infty$). The total kinetic energy in all precipitating 
electrons is $2.6\times 10^{30}$ erg. At peak time, an average of $1.4 \times 10^{36}$ electrons/s precipitated, 
for a total of about $1.1 \times 10^{38}$ electrons.

\emph{Kinetic energy of ejected plasma (bulk motion) $E_{\rm mot}$:} Assuming isothermal plasma, each of TRACE's CCD pixel possesses a flux:
\begin{equation}
	F = f_{195} (T) \cdot EM ,
\label{DN_EM}
\end{equation}
where $F$ is the flux (`data number', in TRACE parlance, normalized to 1 second) in the pixel (CCD dark currents have been subtracted), $EM$ the emission measure observed in that pixel, and $f_{195}(T)$ a known function of the temperature $T$ (the filter response function)
\cite{TRACE1999}. 
As TRACE was observing with only one filter band (195\AA, with one aluminium filter in the FOV), temperature and emission measure cannot be determined without additional information or assumptions.
$f_{195}(T)$ peaks at $T$=1.4 MK. Given a certain flux in a pixel, this temperature yields a lower limit of the emission measure (and hence, the density).
RHESSI spectral fitting (section 5) yields an independent measurement, a temperature of 20.7 MK for the flare kernel. 
Assuming that the ejecta temperature is in the range 1.4-20.8 MK, lower and upper limits can be found for the emission measure,
and in particular, the density.

To determine the total thermal energy and the energy of bulk motion of the ejecta, we need to know the number of particles in the ejecta. 
This was done for the third TRACE image in Figure \ref{TRACEimgs} in the following manner. 
The volume of material in a pixel is $V = $ (pixel area)$ \times l$, where $l$ is the smallest dimension of the ejecta feature being examined. 
Combined with Eq. (\ref{EQ_ne}), and knowing that pixel's emission measure (cf. Eq. \ref{DN_EM}), a density can be calculated.
The average densities $n_{e,ejecta}$ derived in this manner were $5\times 10^9 $cm$^{-3}$ for $T$=1.4 MK and $10^{11} $cm$^{-3}$ for $T$$\sim$10 MK (the temperature were the filter response was lowest).
These densities can be compared with those indicated by the observed decimetric type III bursts in the range 1.2$\times 10^9 < n_e < 1.3 \times 10^{10}$cm$^{-3}$ 
(assuming emission at the second harmonic). 

The shape of the ejecta seen here can be approximated by a truncated cone. The volume of all the ejected material was estimated to be $V_{ejecta}= (1.1\pm0.2) \times 10^{27} $cm$^{3}$.
Hence, $(n_{e,ejecta} \cdot V_{ejecta})$ yields $4.5 \times 10^{36}$ to $1.3 \times 10^{38}$ electrons.
In the flaring kernel, TRACE finds a density of $n_{e,kernel} \approx 3.5(\pm 0.1) \times 10^{11} $cm$^{-3}$, for temperatures
in the range 15-20 MK.


As a self-consistency check, the emission measure of  the brightest EUV region outside the flare kernel (the X-point in Fig. \ref{scenario}, 
fourth image) was determined (assuming an upper limit temperature of 20.8 MK). The result of  $1.8\times 10^{47}$cm$^{-3}$ is below the $EM_{kernel} = 2.2\times 10^{48}$cm$^{-3}$ determined from the RHESSI spectrum for the flare kernel. 
As RHESSI's dynamic range is currently about 10, this means that this region would have been indeed invisible to imaging, even if it were as hot as the flare kernel.

The bulk motion of the ejecta during the impulsive phase of the flare was determined by TRACE difference images (Fig. \ref{TRACEimgs}, two and three), and
was found to be $v = 290\pm{70}$ km/s. It is used to compute the energy of motion of the ejecta, $E_{mot} \approx 0.5 m_p (n_{e,ejecta} \cdot V_{ejecta}) \, v^2$, where $m_p$ is the proton mass. 
This yields a result between $2.0\times 10^{27}$erg and  $1.3\times 10^{29}$erg.

\emph{Thermal energy of flaring kernel $E_{th,kernel}$:} Using the RHESSI-derived values from section 5 and Eq. (\ref{eth}), a value of $6.7 \times 10^{28}$ erg is derived. 
The possible range of values is $3.6-12 \times 10^{28}$ erg.

\emph{Thermal energy of ejected plasma $E_{\rm th,ejecta}$:} We again use $3\, k_BT \,n_e V$, where $T$ is assumed to be between 1.4 and 
20.8 MK, and $n_e$ and $V$ now relate to the ejecta. This yields: $E_{\rm th,ejecta} = 2.6 \times 10^{27}$ to $1.1 \times 10^{30}$ erg.

\emph{Radiated energy from the flaring kernel: $E_{\rm rad,kernel}$:} assuming a temperature of 19.7MK, 
using $EM_{kernel}=2.2 \times 10^{48} $cm$^{-3}$, and integrated between 10:26:05 to 10:27:23 UT, an amount 
of $3.5(\pm1.7) \times 10^{27}$ erg has been radiated as the plasma cooled down.

\emph{Total radiated HXR from precipitating electrons $E_{HXR}$:} integrating the power-law in section 5 
between 10:26:05 to 10:27:23 UT yields a total of $3.2(\pm1.1) \times 10^{23}$ erg.

\begin{table}
\begin{tabular}{|l|c|c|}\hline 
Type &best estimate  & range \\
\hline
$E_{kin,beam}$		& $2.6 \times 10^{30}$  erg	& $1.8 to 3.4 \times 10^{30}$ erg				\\
$E_{th,kernel}$ 	& $6.7 \times 10^{28}$  erg	& 3.6 to 12 $\times 10^{28}$ erg			\\
$E_{th,ejecta}$ 	& $\sim 10^{30}$  erg	& $2.6 \times 10^{26}$ to $1.1 \times 10^{30}$ erg	\\
$E_{mot,ejecta}$ 		& $\sim 10^{29}$ erg		& $2 \times 10^{27}$ to $1.3\times10^{29}$ erg		\\
\hline	
$E_{rad,kernel}$	& $3.5 \times 10^{27}$ erg	& $\pm1.7 \times 10^{27}$ erg				\\
$E_{HXR}$ 		& $3.2 \times 10^{23}$ erg		& $\pm1.1 \times 10^{23}$ erg				\\
\hline
\end{tabular}
\caption {Energy budget of the flare on 2002/02/20 during the impulsive phase (from about 10:26:10 to 10:27:10 UT).} \label{Ebudget}
\end{table}

\section{Conclusions}
The thermal energy content of the hot flare plasma (flare kernel) is considerably less than the energy in the non-thermal electron beam (Table \ref{Ebudget}). 
This is consistent with the standard flare scenario where the energy is first released into non-thermal particles and then converted
into thermal energy. As some of the target may not be heated to high enough temperatures to radiate X-rays, the energy input by the particle beam
can exceed the output visible in soft X-rays.
The ratio of beam energy over thermal energy in the kernel is $\sim$40, much larger than 1, and consistent with results from \inlinecite{deJager1989},
taking into account the fact that they only considered electrons above 25 keV.
A major uncertainty of the energy budget is the source volume, from which several parameters are derived, such as density, mass and energy.
The low-energy cut-off of the non-thermal electron spectrum is the major inaccuracy of the beam energy.

The size of the source was stable in energy and time. The flare kernel contained initially both the thermal and non-thermal electrons.
Later, after the HXR emission above 25 keV ended, the thermal source drifted slowly ($\sim$ 30 km/s) outwards.

If interpreted by reconnection at point A (Fig. \ref{scenario}), the conclusion is that the geometry of energy release and partition was unsymmetrical. 
In the downward jet, not observed by TRACE, the energy was largely transferred to accelerate electrons. 
The proposed scenario also suggests that the accelerated electrons mostly moved downward from the reconnection site or were accelerated only in the downward reconnection jet. If the observed ejecta is interpreted as the other reconnection jet, this upward jet involved less energy, which showed up mostly as heat. 
However, the energy estimate of the latter is less accurate.

RHESSI imaging with improved dynamic range may be able to search for the thin-target emission of energetic electrons in the ejecta. 
Nevertheless, the absence of appreciable decimetric radio emission corroborates the conclusion that acceleration took place mostly below point A and in the downward direction. 

The now available high-resolution RHESSI and TRACE observations allow a more quantitative investigation of flare energies. 
The study of a compact flare yields a detailed scenario (that may not apply to all flares). 
Based on the above interpretations, we conclude that energy partition is not symmetric about the X-point of reconnection. 
Most of the initial energy first appears as energetic electrons in the lower, stationary part, and less than half is manifest in thermal energy and even less in bulk motion of the upper part. 

More flares need to be analyzed to study the influence of the magnetic field geometry and density on energy partition.

\begin{acknowledgements}
We thank the RHESSI software team for continuous encouragements and support,
in particular Gordon Hurford, Andr\'e Csillaghy, Jim McTiernan, and Richard Schwartz for related help
or explanations. Special thanks to Brian Dennis and Ed Schmahl for comments on this paper.
The RHESSI work at ETH Zurich is  supported by the Swiss National
Science Foundation (grant nr. 20-67995.02) and ETH (grant TH-W1/99-2).
\end{acknowledgements}

\end{article}
\end{document}